\input{aipcheck}
\documentclass[
    ,final            
 ,numberedheadings 
  ]
  {aipproc}

\layoutstyle{8x11single}
\usepackage{amsmath,amssymb,amsthm,bm,bbm,color}
\usepackage{graphicx,subfigure,rotating}

\begin{document} 

\title{Random center vortex lines in continuous 3D space-time}



\author{Roman H\"ollwieser}{
address={Department of Physics, New Mexico State University, PO Box 30001, Las
Cruces, NM 88003-8001, USA},
altaddress={Institute of Atomic and Subatomic Physics, Vienna University of
Technology, Operngasse 9, 1040 Vienna, Austria}}
\author{Derar Altarawneh}{
address={Department of Physics, New Mexico State University, PO Box 30001, Las
Cruces, NM 88003-8001, USA},
altaddress={Department of Applied Physics, Tafila Technical University, Tafila , 66110 , Jordan}}
\author{Michael Engelhardt}{
address={Department of Physics, New Mexico State University, PO Box 30001, Las
Cruces, NM 88003-8001, USA}}

\begin{abstract}
We present a model of center vortices, represented by closed random lines in
continuous 2+1- dimensional space-time. These random lines are modeled as being
piece-wise linear and an ensemble is generated by Monte Carlo methods. The
physical space in which the vortex lines are defined is a cuboid with periodic
boundary conditions. Besides moving, growing and shrinking of the vortex configuration, also reconnections are allowed. Our ensemble therefore contains not a fixed, but a variable number of closed vortex lines.
This is expected to be important for realizing the deconfining phase transition.
Using the model, we study both vortex percolation and the potential $V(R)$
between quark and anti-quark as a function of distance $R$  at different vortex
densities, vortex segment lengths, reconnection conditions and at different
temperatures. We have found three deconfinement phase transitions, as a function
of density, as a function of vortex segment length, and as a function of
temperature. The model reproduces the qualitative features of confinement physics seen in $SU(2)$ Yang-Mills theory.\footnote{Presented at 11th Conference on Quark
Confinement and the Hadron Spectrum: ConfinementXI, September, 7-12, 2014, St.
Petersburg State University, Russia, Funded by an Erwin Schr\"odinger Fellowship
of the Austrian Science Fund under Contract No. J3425-N27.}
\end{abstract}

\maketitle

\section{Introduction}
In $D$-dimensional space-time, center vortices are (thickened)
($D-2$)-dimensional chromo-magnetic flux degrees of freedom. The center vortex
picture of the confining vacuum~\cite{'tHooft:1977hy,Yoneya:1978dt,Mack:1978rq,Nielsen:1979xu} assumes that these are the relevant degrees of freedom in the infrared sector of the strong interaction; the center vortices consequently are taken to be weakly coupled and can thus be expected to behave as random lines (for $D = 3$) or random surfaces
(for $D = 4$). The magnetic flux carried by the vortices is quantized in units
which are singled out by the topology of the gauge group, such that the flux is
stable against small local fluctuations of the gauge fields. The vortex model of confinement states that the deconfinement transition is simply a percolation
transition of these chromo-magnetic flux degrees of freedom. 
It is theoretically appealing and was confirmed by a multitude of numerical
calculations, both in lattice Yang-Mills theory and within a corresponding
infrared effective model, see~\cite{Greensite:2003bk} for an overview. Lattice
simulations further indicate that vortices may also be responsible for
topological charge and $\chi$SB~\cite{deForcrand:1999ms,Alexandrou:1999vx,Engelhardt:2002qs,Bornyakov:2007fz,Hollwieser:2008tq,Hollwieser:2011uj,Hollwieser:2013xja},
and thus unify all non-perturbative phenomena in a common framework.  

We introduce a model of random flux lines in $D=2+1$ space-time dimensions. 
The lines are composed of straight segments connecting nodes randomly distributed 
in three-dimensional space. Allowance is made for nodes moving as well as being
added or deleted from the configurations during Monte Carlo updates.
Furthermore, Monte Carlo updates disconnecting and fusing vortex lines, {\it
i.e.} reconnection updates were implemented. Given that the deconfining phase
transition is a percolation transition, such processes play a crucial role in
the vortex picture.  The model has been formulated in a finite volume, with
periodic boundary conditions, which will allow for a study of finite
temperatures (via changes in the temporal extent of the volume). The resulting
vortex ensemble will be used, in particular, to measure the string tension and
its behavior as a function of temperature, with a view towards detecting the
high-temperature deconfining phase transition. 

\section{The Model}\label{sec:model}
The physical space in which the vortex lines are defined is a cuboid
$L_S^2\times L_T$ with "spatial" extent $L_S$, "temporal" extent $L_T$ and periodic
boundary conditions in all directions. The random lines are modeled as being
piece-wise linear between nodes, with vortex length $L$ between two nodes restricted to a
certain range $L_{min}<L<L_{max}$. This range in some sense sets the scale of
the model; for practical reasons we choose a scale of $L\approx1$, {\it i.e.}
$L_{min}=0.3$ and $L_{max}=1.7$ in these dimensionless units. Within this paper
we use volumes with $L_S=16$, where finite size effects are under control, and
varying time lengths $L_T$. The variation of the vortex lengths range will also
be examined. An ensemble is generated by Monte Carlo methods, starting with a
random initial configuration. A Metropolis algorithm is applied to all
updates using the action $S= \alpha L + \gamma \varphi^2,\label{eq:act}$
with action parameters $\alpha$ and $\gamma$ for the vortex length $L$ and the
vortex angle $\varphi$ between two adjacent segments respectively. At a given (current) node the vortex length $L$ is defined to be the distance to the previous node and the vortex angle is the angle between the oriented vectors of the vortex lines connecting the previous, current and next nodes. When two vortices approach each other, they can reconnect or separate at a bottleneck. The ensemble therefore will contain not a fixed, but a variable number of closed vortex lines or "vortex clusters". This is expected to be important for realizing the deconfining phase transition. 

Move, add and delete updates are applied to the vortex nodes via the Metropolis algorithm, {\it i.e.} the difference of the action of the affected
nodes before and after the update determines the probability of the update to be
accepted. The move update moves the current node by a random vector of
maximal length $r_m=4L_{min}$, it affects the action of three nodes, the node
itself and its neighbors. The add update adds a node at a random position within a radius
$r_a=3L_{min}$ around the midpoint between the current and the next node. 
The action before the update is given by the sum of the action at the current and the next node, while the action after the update is the sum of the action at the current, the new and the next node. Deleting the current node, on the other hand, affects three nodes before the update and only two nodes after the update. Therefore the
probability for the add update is in general much smaller than for the
delete update and the vortex structure would soon vanish if both updates were
tried equally often. Hence, the update strategy is randomized to move a node in two out
of three cases ($66\%$), and apply the add update about five times more often
than the delete update ($28\%$ vs. $6\%$). If the current node is not deleted, all nodes around the current node are considered for reconnections. The reconnection update causes
the cancellation of two close, nearly parallel vortex lines and reconnection of the
involved nodes with new vortex lines. The reconnection update is also subjected to the Metropolis algorithm, considering the action of the four nodes involved. All updates resulting in vortex lengths $L$ out of the range $L_{min}<L<L_{max}$ are rejected. 
Finally, a density parameter $d$ is introduced, restricting the number of nodes in a certain volume. The add update is rejected if the number of nodes within a $3\times3\times3$ volume around the new node exceeds the density parameter $d$. 

\section{Measurements and Observables}\label{sec:measo}
For our simulations we start random vortex line configurations and do $10^4$
thermalization sweeps through the whole configuration, {\it i.e.} updating all
single nodes within each sweep. After that $2-5*10^5$ measurements of the
average action per node, the actual density, average vortex length and angle and
Wilson loops were measured and a vortex cluster analysis was performed every $n_s=10$ Monte Carlo sweeps. 
The vortex cluster routine counts the number of closed vortex clusters, the
number of vortex nodes/lines for each cluster, the cluster size or maximal
extent of each cluster and the number of clusters winding around the time
dimension. Taking into account the periodic boundary conditions, the maximal
cluster size is given by $s_m=L_S^2/2+L_T^2/4$.
The Wilson loop $W(R,T)$ is a common observable in lattice gauge field theory.
It is a closed (rectangular) loop in space and time and can be interpreted as
the creation of a quark--anti-quark pair with a certain spatial separation $R$
evolving for some time $T$ and annihilating again. In our continuous space model
we are not restricted to rectangular loops, but they are more convenient for actual
calculations. The Wilson loop is calculated by counting the number $n_p$ of vortex
lines piercing through a plane of extent $R\times T$, resulting in
$W(R,T)=(-1)^{n_p}$. 
The expectation values of the time-like Wilson loops
$\langle W(R,T)\rangle$ give the quark--anti-quark potential 
$V(R)\propto -\underset{T\rightarrow\infty}{lim}\;log\langle W(R,T)\rangle /T$. 
To extract the string tension $\sigma$ of the system an ansatz $V(R)=\sigma
R+C/R+V_0$ is fitted to the potentials. The spatial string tension $\sigma_s$
is obtained from spatial Wilson loops.

\section{Results and Discussion}\label{sec:res}

\begin{figure}[h]
	\centering
	c)\includegraphics[width=.47\linewidth]{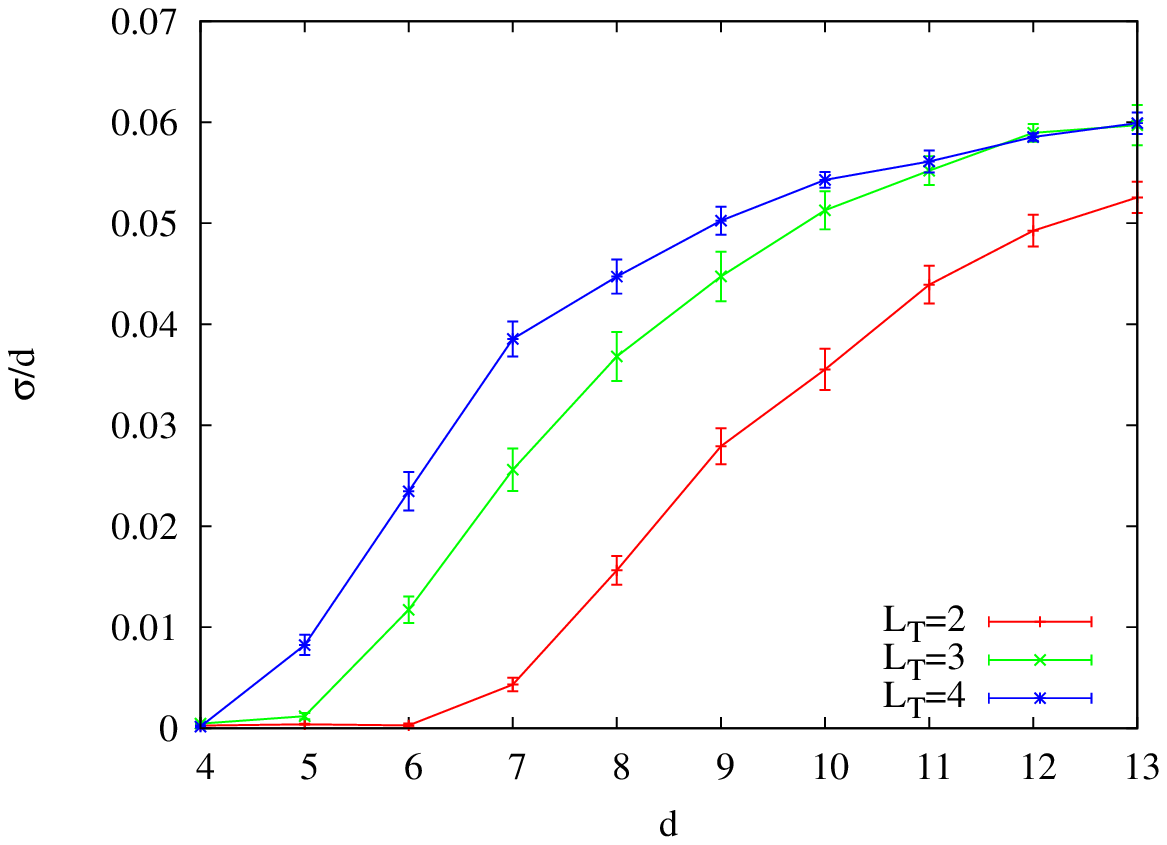}
	d)\includegraphics[width=.47\linewidth]{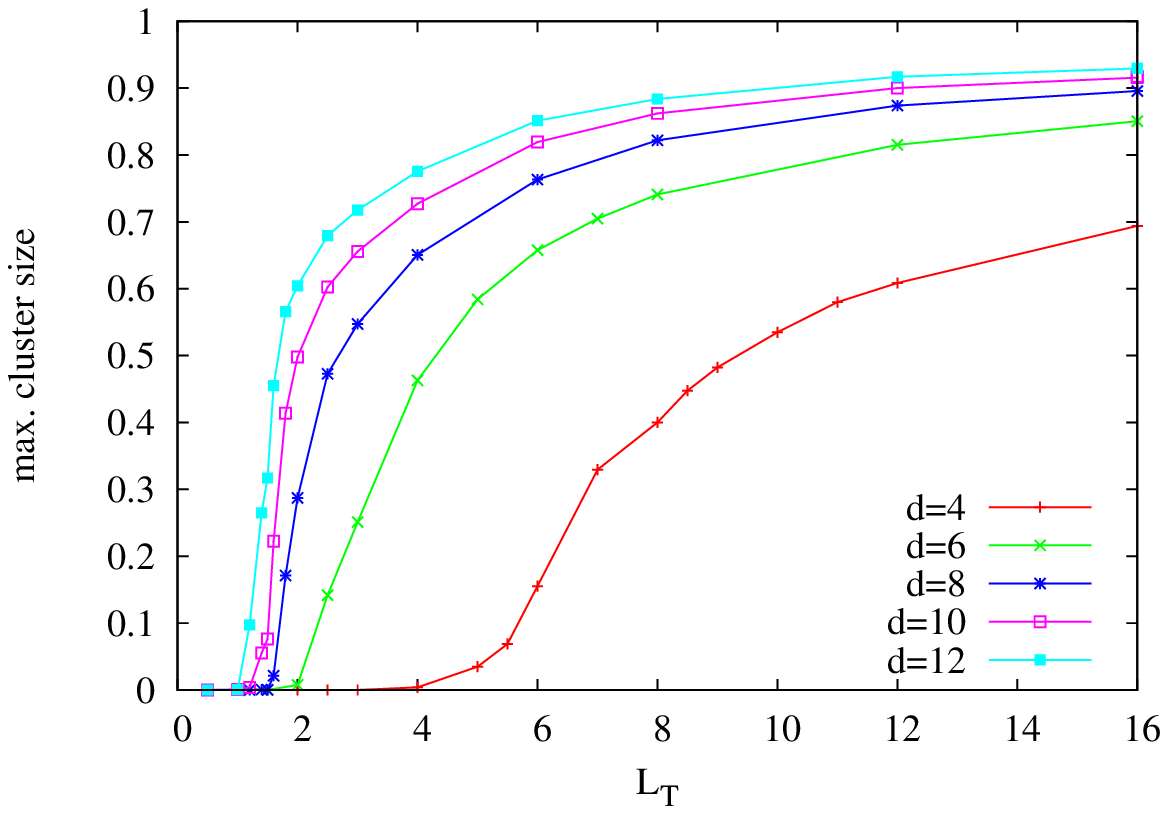}
	\caption{Results for $16^2\times L_T$ volumes with vortex lengths
	$L_{max}=1.7$ and $L_{min}=0.3$: a) Quark--anti-quark potentials for density
	cutoff $d=4$, b) string tension $\sigma$ and d) maximum cluster size vs.
	time extent $L_T$, note the good correlation between the two, and c) string
	tension $\sigma/d$ vs. density cutoff $d$.}
\label{fig:d4}
\end{figure}

In Fig.~\ref{fig:d4} we show the the potential $V(r)$ between quark
and anti-quark and the string tension $\sigma$ as well as the
maximal cluster size as a function of inverse temperature $L_T$ and density
cutoff $d$. 
We observe a phase transition as a function of the inverse temperature $L_T$. 
The quark--anti-quark potential shown in Fig.~\ref{fig:d4}a is
still flat (asymptotically) for $L_T=4$ while we may see some linearly rising
behavior for $L_T=9$. In between the potentials do not show an exactly linear
behavior, and the determination of the string tension is somewhat ambiguous, but
the bending  is of course caused by the finite spatial extent of the physical
volume. In the $d=4$ case there is no sharp but a rather smooth
transition, which is caused by the small density cutoff.
The transition gets sharper and the transition temperature increases with the
density cutoff $d$. We locate the phase transition for $d=6$ between $L_T=2-4$, for $d=8$ between $L_T=1.6-2.5$, for $d=10$ between $L_T=1.5-2$ and for $d=12$ between
$L_T=1.2-1.5$. Further, we notice that Fig.~\ref{fig:d4}b and d show a perfect agreement between confinement (string tension) and  percolation (maximal cluster size) transition. 

\begin{figure}[h]
	\centering
	a)\includegraphics[width=.47\linewidth]{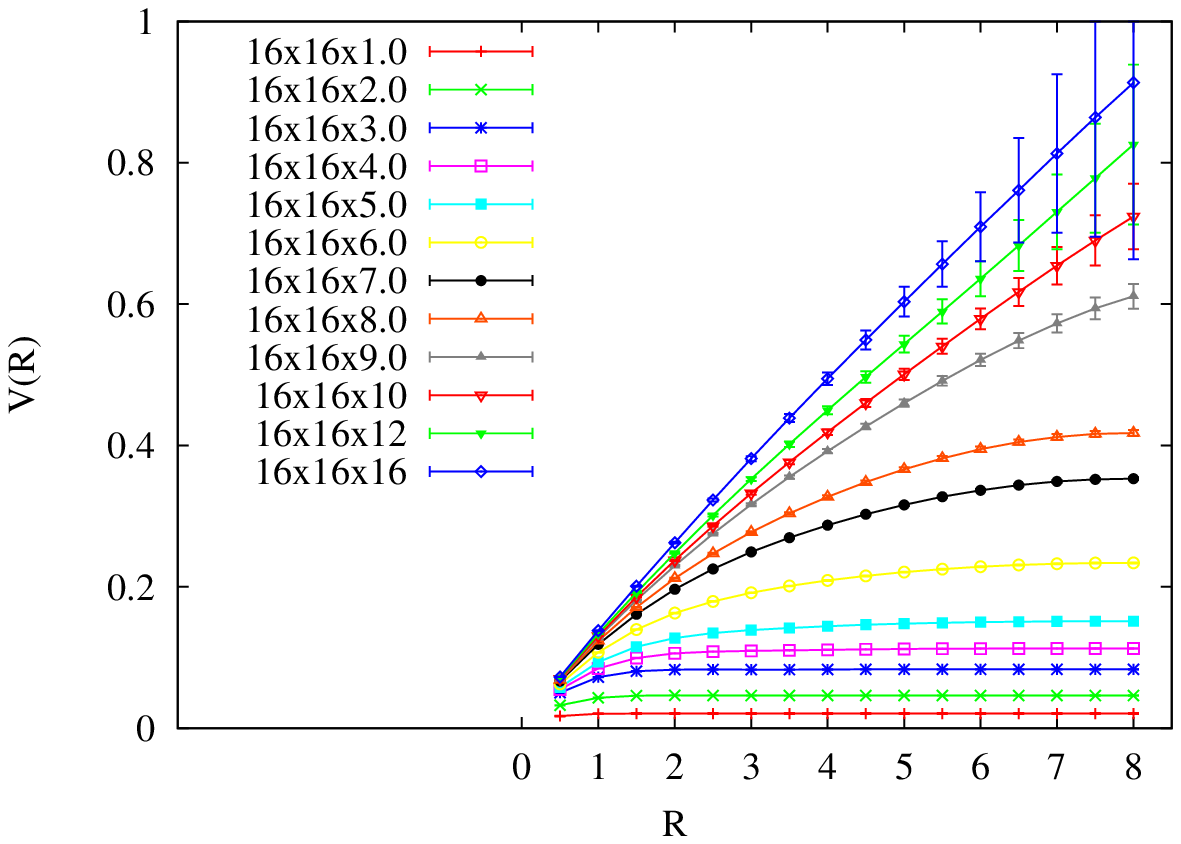}
	b)\includegraphics[width=.47\linewidth]{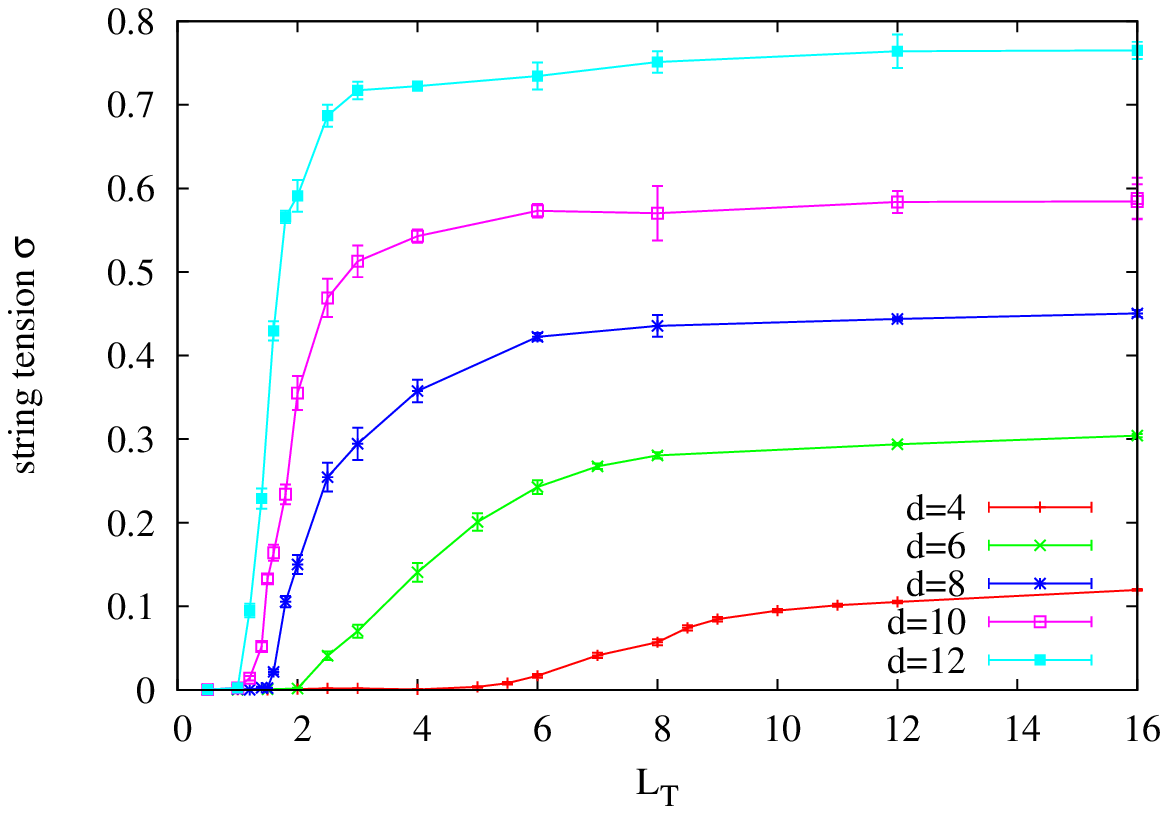}
\end{figure}

In Fig.~\ref{fig:d4}c we show the string tensions $\sigma/d$ for different density cutoffs $d=4-13$ for three different physical volumes $16^2\times L_T$ with $L_T=2-4$, $L_{max}=1.7$ and $L_{min}=0.3$. 
We observe a confinement phase
transition with respect to the vortex density cutoff $d$. At $d=4$ all
configurations are in the deconfined phase, the $L_T=4$ configurations
immediately start to confine with increasing $d$, whereas $L_T=3$ and $L_T=2$
configurations start the phase transition at $d=5$ and $d=6$, respectively. 

\begin{figure}[h]
	\centering
	a)\includegraphics[width=.47\linewidth]{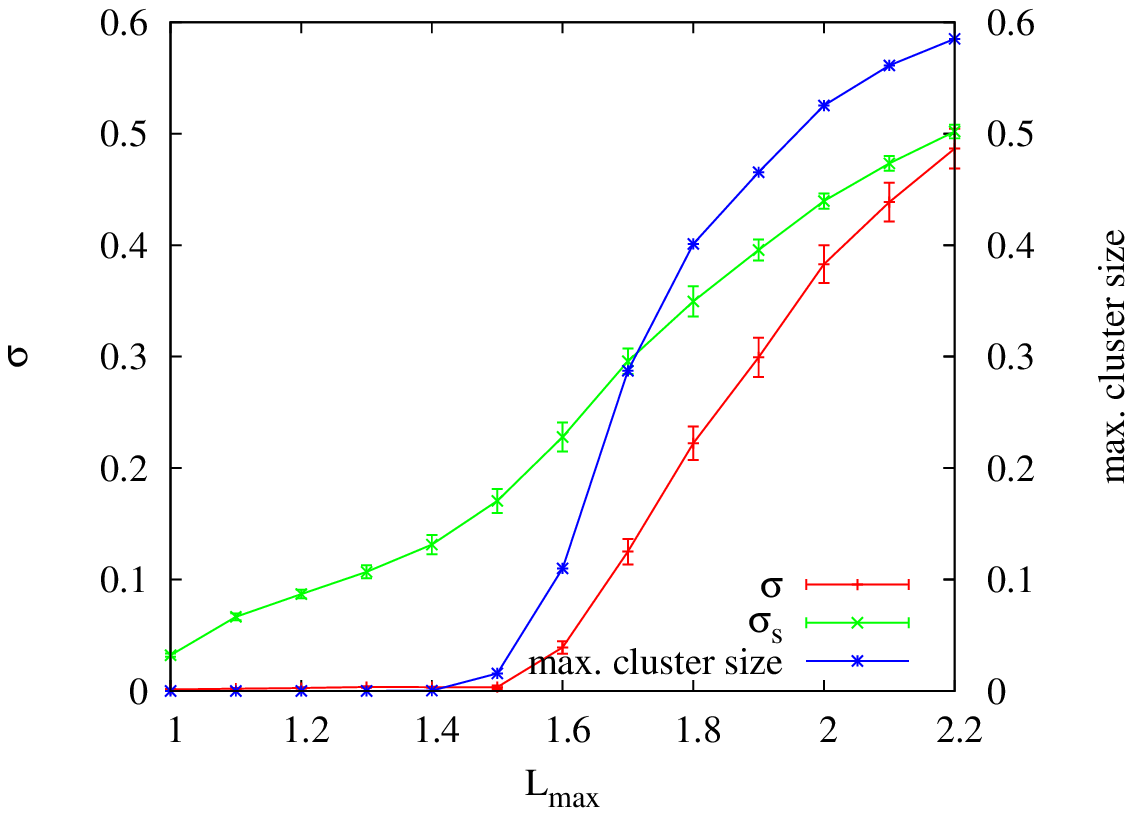}
	c)\includegraphics[width=.47\linewidth]{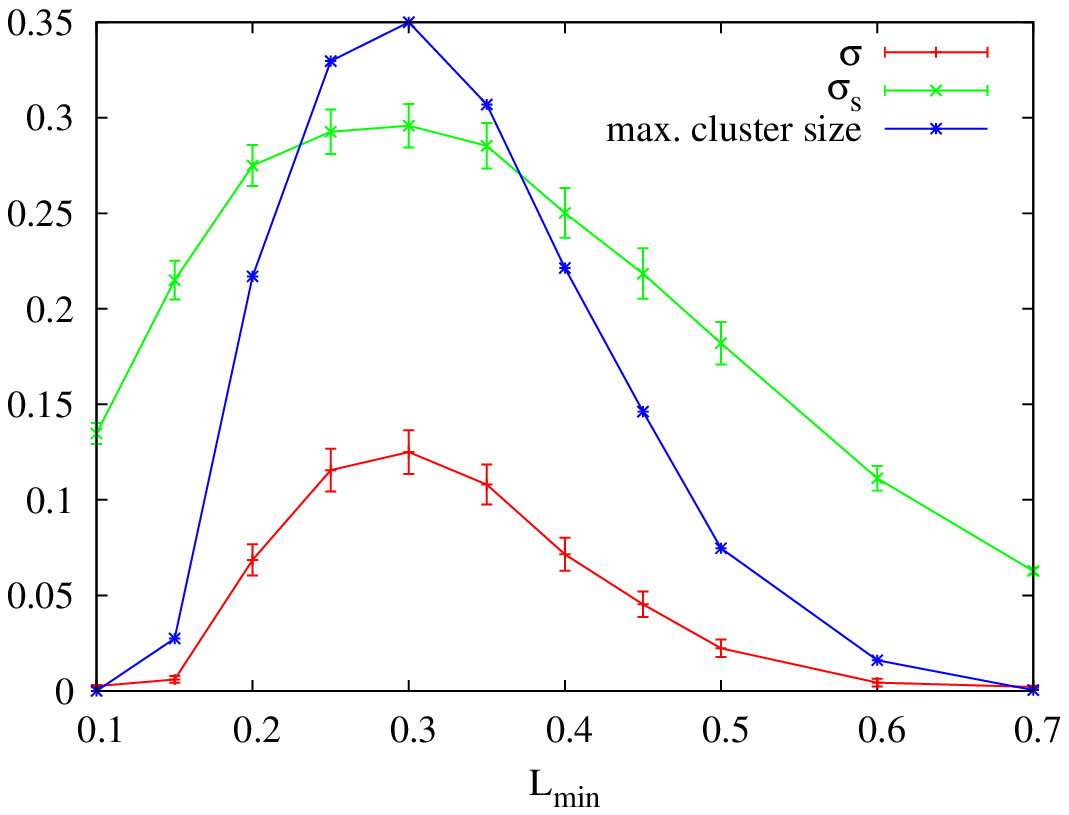}
	\caption{String tensions $\sigma$, $\sigma_s$ and maximal cluster size
	for $16^2\times2$ volumes and density cutoff $d=8$ vs. vortex lengths a) $L_{max}$ and b) $L_{min}$.}\label{fig:lmax}
\end{figure}

Another phase transition is observed for different maximal vortex segment
lengths $L_{max}=1.0-2.2$ for a physical volume $16^2\times2$ and a density
cutoff $d=8$, with $L_{min}=0.3$. In Fig.~\ref{fig:lmax}a we show the string tensions $\sigma$ and $\sigma_s$ as well as maximal cluster size versus the different vortex lengths $L_{max}$. We observe a well defined phase transition starting at $L_{max}=1.5$, with no percolation and string tension $\sigma$ below that threshold and percolation resp. confinement above. 
Restricting the vortex line length to small values leads to small, separated
vortex clusters which cannot reconnect or percolate and confinement is lost.

The interpretation of the phase space with respect to the minimal vortex length
$L_{min}$ is going to be more complex than in all other cases, since $L_{min}$
defines some sort of minimal length scale and we use it to restrict the minimal
length of a vortex segment itself, the move radius $r_m=4L_{min}$, the add
radius $r_a=3L_{min}$ and the reconnection length $r_r=L_{min}$. That means that
if we choose a small $L_{min}$, the vortex clusters will not spread out very
fast and reconnections are strongly suppressed. On the other hand a large
$L_{min}$ will lead to large random fluctuations and a huge number of random
recombinations leading to a very chaotic system. Both cases do not seem to realize
the physical systems we want to study and our test configurations in
$16^2\times2$ volumes with vortex density $d=8$, maximal vortex length
$L_{max}=1.7$ and varying $L_{min}$ seem to confirm these expectations. This
setup happens to lie exactly at the critical value for the parameters of
all the above mentioned phase transitions, {\it i.e.} finite temperature,
vortex density and maximal vortex length $L_{max}$. In Fig.~\ref{fig:lmax}b we
plot the string tensions $\sigma$, $\sigma_s$ and maximal cluster size 
versus $L_{min}=0.1-0.7$ in steps of $0.05$. We
note deconfined phases for both, small and large $L_{min}$, on the one hand
because of rather static and on the other hand because of very chaotic configurations,
which both do not seem to show good percolation properties. At $L_{min}=0.3$
however, we find a common maximum for string tensions, maximal cluster size and
average vortex density and simultaneously the average action shows a minimum.
This finally motivates our initial choice for $L_{min}=0.3$, which seems to give
rather stable configurations which allowed reliable studies of the above phase
transitions. 

\begin{figure}[h]
	\centering
	a)\includegraphics[width=.24\linewidth]{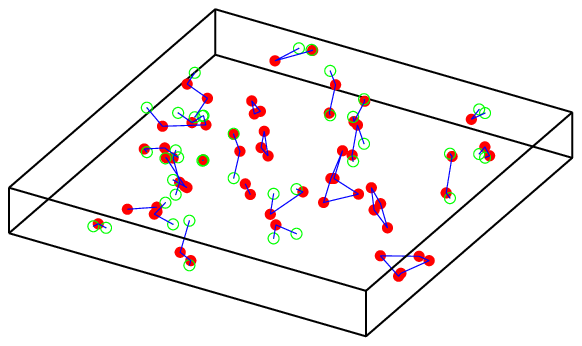}
b)\includegraphics[width=.24\linewidth]{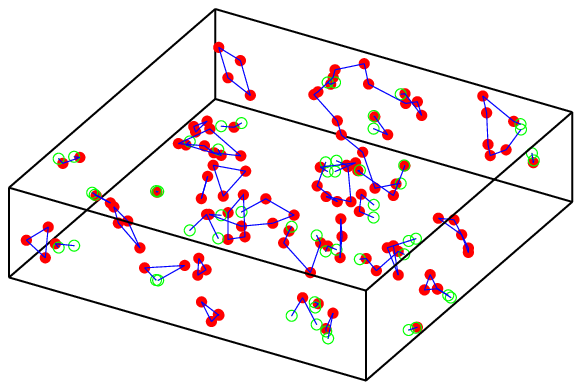}
c)\includegraphics[width=.24\linewidth]{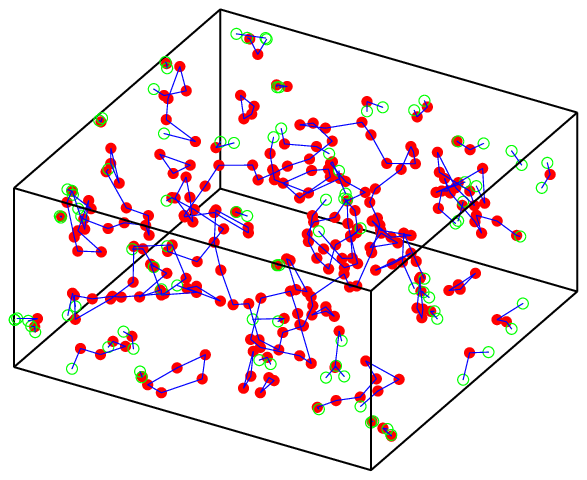}
d)\includegraphics[width=.24\linewidth]{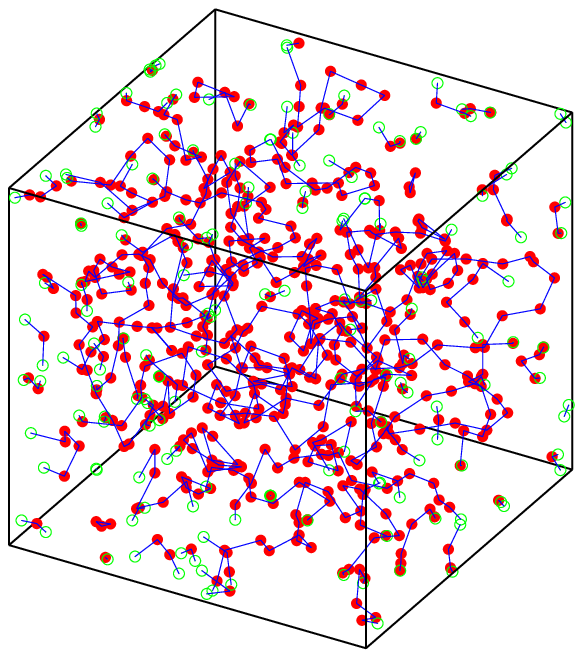}
	\caption{$16^2\times$ a) $2$, b) $4$, c) $8$ and d) $16$ configurations, density cutoff $d=4$, vortex lengths $L_{max}=1.7, L_{min}=0.3$.}
	\label{fig:conf3D}
\end{figure}

In Fig.~\ref{fig:conf3D} we show sample configurations for various temperatures and
density cutoff $d=4$. While for $L_T=2$ and $4$ (Fig.~\ref{fig:conf3D}a and b)
we see many small vortex clusters,  we observe already one big cluster extending
over the whole physical volume together with some small clusters for $L_T=8$
(Fig.~\ref{fig:conf3D}c) while for $L_T=16$ (zero temperature,
Fig.~\ref{fig:conf3D}d) it seems as if almost all nodes were connected,
representing the percolation property.

\section{Conclusions and Outlook}\label{sec:concl}

We introduced a $2+1$ dimensional center vortex model of the Yang-Mills vacuum.
The vortices are represented by closed random lines which are modeled as being
piece-wise linear and an ensemble is generated by Monte Carlo methods. The
physical space in which the vortex lines are defined is a cuboid with periodic
boundary conditions. The idea was to simulate continuous space-time,
hence in contrast to existing discrete vortex models, the vortices are allowed
to move freely and we have translational and rotational symmetry in our system.
We further let vortex configurations grow and shrink and also reconnections are
allowed, {\it i.e.} vortex lines may fuse or disconnect. Our ensemble therefore
contains not a fixed, but a variable number of closed vortex lines. The
reconnections are in fact the important ingredient for modeling a percolating
system which is supposed to drive the confinement phase transition. All updates
(move, add, delete, reconnect) are subjected to a Metropolis algorithm driven by
an action depending on vortex segment length and the angle between two vortex
segments, {\it i.e.} a length and curvature part. After fine-tuning all
necessary parameters, we use the model to study both vortex percolation and the
potential $V(R)$ between quark and anti-quark as a function of distance $R$  at
different vortex densities, vortex segment lengths, reconnection conditions and
at different temperatures (by varying the temporal extent of the physical volume).

We have found three confinement phase transitions, as a function of density, as a function of vortex segment length, and as a function of temperature. The confinement phase
transitions are in perfect agreement with the percolation properties of the
vortex configuration. For small vortex densities and vortex segment lengths the
configuration consists of small, independent vortex clusters and for high
temperatures vortex clusters prefer to wind around the (short) temporal extent
of the volume and therefore these configurations are not percolating. The
quark--anti-quark potentials in this case show no linearly rising behavior, {\it
i.e.} no string tension is measured and the system is in the deconfined phase.
Once we allow higher vortex densities, longer vortex segment lengths or temporal
extent, {\it i.e.} lower temperature, the vortex configurations start to
percolate and small clusters reconnect to mainly one big vortex cluster filling
the whole volume. In this regime we measure a finite string tension, {\it i.e.}
linearly rising quark--anti-quark potentials, hence the vortices confine quarks
and anti-quarks. We also study the influence of reconnection parameters, which are obviously very crucial for the percolation properties. These parameters have to be fine-tuned very carefully in order to find the above phase transitions. 

We plan the expansion of the model to four dimensions, where some
sort of random triangulation of vortex surfaces is needed, which then allows
to study topological properties in addition to the confinement transition.
The complex parameter space of the three dimensional case definitely indicates
the difficulty of this task, but still, the model constructed in this paper certainly reproduces the qualitative features of confinement physics seen in $SU(2)$ Yang-Mills theory.

\begin{theacknowledgments}
This research was supported by the U.S. DOE through the grant DE-FG02-96ER40965
(D.A.,M.E.) and the Erwin Schr\"odinger Fellowship program of the Austrian
Science Fund FWF (``Fonds zur F\"orderung der wissenschaftlichen
Forschung'') under Contract No. J3425-N27 (R.H.).
\end{theacknowledgments}

\bibliographystyle{aipproc}
\bibliography{literatur}

\end{document}